\documentclass[sigplan,screen]{acmart} 

\AtBeginDocument{%
  }

\usepackage{amsfonts}
\usepackage{graphicx}
\usepackage{textcomp}
\usepackage{balance}
\usepackage{fancyhdr}
\usepackage{enumitem}
\usepackage{nicefrac}
\usepackage[linesnumbered, ruled, vlined]{algorithm2e}
\usepackage[most]{tcolorbox} 
\usepackage{marginnote}

\definecolor{olivegreen}{rgb}{0, 0.6, 0}
\definecolor{redorange}{HTML}{FF5349}
\definecolor{blue(ncs)}{rgb}{0.0, 0.53, 0.74}
\definecolor{navy}{HTML}{273BE2}
\definecolor{black}{HTML}{000000}
\definecolor{white}{HTML}{ffffff}
\definecolor{greenyellow}{HTML}{DDE576}

\usepackage{xspace}

\usepackage{mathtools, nccmath}

\usepackage{multicol}
\usepackage{multirow}
\usepackage{makecell}
\usepackage{booktabs}

\usepackage[noabbrev, capitalize]{cleveref}

\usepackage{comment}

\usepackage{bookmark}

\usepackage{nicefrac}

\newcommand{\thiswork}{HILOS\xspace}

\newcommand{\rev}[1]{{#1}}

\newcommand{\JL}[1]{{\color{cyan}[\textbf{\sc JLee}: \textit{#1}]}}
\newcommand{\JW}[1]{{\color{orange}[\textbf{\sc JJung}: \textit{#1}]}}
\newcommand{\JY}[1]{{\color{blue(ncs)}[\textbf{\sc JSong}: \textit{#1}]}}
\newcommand{\HS}[1]{{\color{magenta}[\textbf{\sc HJang}: \textit{#1}]}}
\newcommand{\CS}[1]{{\color{navy}[\textbf{\sc CShin}: \textit{#1}]}}
\newcommand{\SN}[1]{{\color{olive}[\textbf{\sc SNoh}: \textit{#1}]}}

\newcommand{\js}[1]{{\color{pred}{#1}}}
\definecolor{pred}{rgb}{0.7843, 0.0039, 0.3137}

\ifdefined\final
\renewcommand{\JL}[1]{}
\renewcommand{\JW}[1]{}
\renewcommand{\JY}[1]{}
\renewcommand{\HS}[1]{}
\renewcommand{\CS}[1]{}
\renewcommand{\SN}[1]{}
\renewcommand{\js}[1]{}
\fi

\newcommand{\baseline}{offloading-based batched inference\xspace}

\newcommand{\ans}{attention near storage\xspace}
\newcommand{\Ans}{Attention near storage\xspace}
\newcommand{\ANS}{Attention Near Storage\xspace}

\newcommand{\wb}{delayed KV cache writeback\xspace}

\newcommand{\WB}{Delayed KV Cache Writeback\xspace}

\newcommand{\xcache}{X-cache\xspace}

\usepackage{tikz}
\newcommand*\circled[1]{\tikz[baseline=(char.base)]{
            \node[shape=circle,draw,inner sep=0.4pt] (char) {#1};}}

\newcommand*\bcircled[1]{\tikz[baseline=(char.base)]{
            \node[shape=circle,draw,inner sep=0.4pt, fill=black, text=white] (char) {#1};}}

\crefname{section}{\S}{\S}

\copyrightyear{2026}
\acmYear{2026}
\setcopyright{cc}
\setcctype{by}
\acmConference[ASPLOS '26]{Proceedings of the 31st ACM International Conference on Architectural Support for Programming Languages and Operating Systems, Volume 2}{March 22--26, 2026}{Pittsburgh, PA, USA}
\acmBooktitle{Proceedings of the 31st ACM International Conference on Architectural Support for Programming Languages and Operating Systems, Volume 2 (ASPLOS '26), March 22--26, 2026, Pittsburgh, PA, USA}
\acmDOI{10.1145/3779212.3790119}
\acmISBN{979-8-4007-2359-9/2026/03}

\author{Hongsun Jang}
\orcid{0000-0003-4291-6124}
\affiliation{%
  \institution{Seoul National University}
  \city{Seoul}
  \country{South Korea}
}
\email{hongsun.jang@snu.ac.kr}

\author{Jaeyong Song}
\orcid{0000-0001-9976-7487}
\affiliation{%
  \institution{Seoul National University}
  \city{Seoul}
  \country{South Korea}
}
\email{jaeyong.song@snu.ac.kr}

\author{Changmin Shin}
\orcid{0009-0004-8242-8330}
\affiliation{%
  \institution{Seoul National University}
  \city{Seoul}
  \country{South Korea}
}
\email{scm8432@snu.ac.kr}

\author{Si Ung Noh}
\orcid{0009-0000-2228-2131}
\affiliation{%
  \institution{Seoul National University}
  \city{Seoul}
  \country{South Korea}
}
\email{siung98@snu.ac.kr}

\author{Jaewon Jung}
\orcid{0000-0003-0770-9277}
\affiliation{%
  \institution{Seoul National University}
  \city{Seoul}
  \country{South Korea}
}
\email{jungjaewon@snu.ac.kr}

\author{Jisung Park}
\orcid{0000-0002-1826-9003}
\affiliation{%
  \institution{POSTECH}
  \city{Pohang}
  \country{South Korea}
}
\email{jisung.park@postech.ac.kr}

\author{Jinho Lee}
\orcid{0000-0003-4010-6611}
\affiliation{%
  \institution{Seoul National University}
  \city{Seoul}
  \country{South Korea}
}
\email{leejinho@snu.ac.kr}

\renewcommand{\shortauthors}{Hongsun Jang et al.}

\begin{document}

\title{
A Cost-Effective Near-Storage Processing Solution for Offline Inference of Long-Context LLMs
}


\begin{abstract}
The computational and memory demands of large language models for generative inference present significant challenges for practical deployment.
One promising solution targeting offline inference is \textit{offloading-based batched inference}, which extends the GPU's memory hierarchy with host memory and storage.
However, it often suffers from substantial I/O overhead, primarily due to the large KV cache sizes that scale with batch size and context window length.

In this paper, we introduce \emph{\thiswork}, a framework that boosts offline inference throughput using near-storage processing.
The core of \thiswork is \emph{\ans}, which offloads memory-intensive attention operations to near-storage accelerators, reducing traffic across the system interconnect.
Building on \ans, \thiswork incorporates three additional optimizations. 
First, \emph{cooperative \xcache} minimizes KV cache I/O by exploiting available host resources after offloading.
Second, \emph{\wb} hides storage write latency and mitigates storage write amplification.
Finally, a \emph{memory-efficient attention accelerator} sustains high throughput for long sequences 
within the resource constraints of NSP devices.
We implemented and evaluated \thiswork on a real system equipped with 16 SmartSSDs.
Compared to state-of-the-art offloading-based inference frameworks, \thiswork achieves up to 7.86$\times$ throughput while reducing energy consumption by up to 85\%.
The source code for \thiswork is available at \url{https://github.com/hongsunjang/HILOS}.

\end{abstract}

\ccsdesc[500]{Computer systems organization~Heterogeneous (hybrid) systems}
\ccsdesc[500]{Computing methodologies~Neural networks}


\keywords{
Near-Storage Processing, Transformer-Based Generative Model, FPGA, Batch Inference
}

\maketitle


\section{Introduction}
\label{sec:intro}

Transformer-based large language models (LLMs)~\cite{attention, gpt2, gpt3, gpt4, llama, opt} are at the forefront of modern artificial intelligence and are expected to greatly expand human productivity~\cite{auto_create1,github_copilot,llm_cloud_incident}. 
Consequently, the efficient support of generative inference for these models~\cite{flexgen, vllm, deepspeedinf, huggingface, petals} has become a critical challenge in modern computing systems. 

While current LLM systems often focus on online inference~\cite{vllm, distserve, orca, splitwise}, offline inference is also critical for various scenarios~\cite{flexgen, deepspeedinf, instattn}.
In contrast to online inference that prioritizes low latency (i.e., low TTFT and tight TPOT constraint), 
offline inference generally allows longer sequences~\cite{openbookqa, chang2024booookscore}, tolerating higher latency. 
It is especially useful for applications such as benchmarking~\cite{liang2023holistic, NEURIPS2023_89e44582, NEURIPS2023_91f18a12, guo2023gpt4graph} and information extraction~\cite{narayan2018don, pu2023summarization, chang2024booookscore, goyal2022news, pang-etal-2023-long}.




A significant challenge in offline inference is the massive memory footprint it requires.
The model parameters and the key-value (KV) cache~\cite{scaletrans} for large models often exceed the memory capacity of a single GPU, especially with the long sequences common in offline tasks.
In such scenarios, equipping costly servers with multiple GPUs to meet increasing memory requirements is often cost-prohibitive.
Furthermore, this approach is less economical because the decoding phase of LLM inference is fundamentally memory-bound, leaving expensive GPU compute units severely underutilized~\cite{splitwise,distserve,attacc,neupims}.

A popular direction is the \emph{\baseline} approach~\cite{flexgen, infinigen, deepspeedinf}. 
It generally offloads model weights and KV caches to host memory, spilling over to secondary storage if the host memory capacity is exceeded.
By batching multiple tokens, it enables repeated reuse of model weights for increased throughput. 
However, an unsurprising consequence is the I/O overhead resulting from KV cache transfers, which is exacerbated by larger batch sizes and longer context lengths~\cite{vllm, infinigen, h2o, flexgen}.
Our study (\cref{sec:motiv1}) shows that KV cache I/O in the \baseline accounts for over 60\% of inference time.


Given these limitations, we advocate the use of near-storage processing (NSP) as an attractive opportunity to address this bottleneck.
NSP offers several key advantages for KV cache storage: (1) abundant storage capacity for terabyte-scale KV caches, (2) high aggregate bandwidth that scales with the number of devices, (3) reduced host I/O overhead. 

In this paper, we propose \emph{\thiswork}, an NSP-based
framework for high-throughput offline LLM inference on a real system.
The core of \thiswork is \textit{\ans}, which processes KV cache-related operations 
within a custom accelerator in each NSP device.
With \ans, only the computed results are sent back to the host, significantly reducing the total I/O overhead. 
While offloading of attention operations has been explored on other platforms in previous work~\cite{neupims, instattn, duplex, attacc}, \thiswork is the first to implement this on off-the-shelf NSP products~\cite{smartssd, falcon}, addressing several practical challenges as follows.


First, offloading KV cache operations to NSP devices shifts the performance bottleneck, causing the internal storage I/O to dominate end-to-end latency.
Based on the observation that this offloading leaves host resources (i.e., the GPU) underutilized, we introduce \emph{cooperative \xcache}, a technique that stores pre-projection activations ($X$) and utilizes remaining host resources to further reduce the internal traffic.

Second, writing the new KV entries back to storage incurs significant latency due to small, inefficient writes.
We propose \emph{\wb}, which buffers new KV entries in host memory and provides partially computed results to the accelerator. 
This not only puts the overhead off the critical path but also allows for efficient updates. 

Third, the storage-side resources are often severely constrained, which makes it challenging to provide enough computational throughput for long-sequence attentions.
To address this challenge, we design a novel architecture that efficiently utilizes off-chip memory for NSP environments. 


We evaluate \thiswork on an end-to-end real-system prototype built using commercial SmartSSDs~\cite{smartssd}.
Our prototype is a ready-to-use system fully integrated into a PyTorch offloading-based LLM inference framework~\cite{flexgen}.
We demonstrate that \thiswork cost-effectively supports 175 billion-parameter LLMs with a 128K context length on a single A100 GPU, without requiring any custom modifications to the existing hardware or model architecture.

\begin{samepage}
Our contributions can be summarized as follows: 
\begin{itemize}[leftmargin=*, noitemsep, topsep=0pt]
    \item We introduce \thiswork, the first NSP-offloading system with an off-the-shelf product that can efficiently support long-context (>100K) LLMs.
    \item We propose three key designs to address the limitations of the naive NSP-offloading system: 
    (1)~cooperative \xcache, which allows for cooperative utilization of host resources alongside NSP devices.
    (2)~\wb, which addresses the inefficient host-storage data transfer for new KV entries, and
    (3)~an efficient NSP accelerator, which enables high scalability for long-context LLMs.
    \item We implement \thiswork on a real system using commodity NSP devices (SmartSSDs~\cite{smartssd}) and integrate it into a PyTorch-based LLM inference framework.
    Our evaluation demonstrates up to a 7.86$\times$ throughput improvement over conventional SSD-based solutions. 
\end{itemize}
\end{samepage}

\section{Background}
\label{sec:background}

\subsection{KV Cache in Transformer-based LLM Inference}
\label{sec:background1}

LLM generative inference generates an output sequence of length $n$ from an input context $X$ of length $s$.
The process operates in two phases: \emph{prefill} and \emph{decoding}.
After the prefill phase processes the entire input context to compute an initial \emph{key-value (KV) cache}~\cite{scaletrans} and the first token,  
the decoding phase generates the remaining $n-1$ tokens auto-regressively.
At each decoding step, the KV cache is reused to attend to all previous tokens to avoid redundant computation
~\cite{scaletrans,flexgen,infinigen,vllm}.
In each transformer layer, the input $X$ is projected using learnable weight matrices $W_Q$, $W_K$, and $W_V$ to generate the Query ($Q$), Key ($K$), and Value ($V$) matrices. 
The attention operation is then computed as shown in the equations below, where $d$ denotes the hidden dimension size per head:
\begin{align}
    Q = X\cdot W_Q; K = X\cdot W_K; V = X\cdot W_V
    \label{eq:qkv} \\
    Attention(Q,K,V) = softmax\ \!\bigl(\tfrac{QK^T}{\sqrt{d}}\bigr)V
    \label{eq:attention}
\end{align}
\subsection{Offloading-based Batched Inference}
\label{sec:background2}

The substantial memory required for LLM inference often exceeds the memory capacity of multiple GPUs.
For offline scenarios, offloading-based batched inference~\cite{flexgen,instattn,deepspeedinf,infinigen} provides a low-cost alternative by extending GPU memory with host resources like CPU memory and SSDs.

\begin{figure}[t]
    \centering
    \includegraphics[width=\columnwidth]{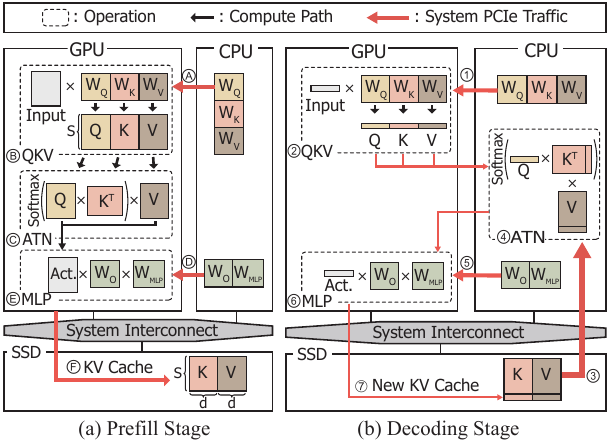} 
    \Description{Background} 
    \caption{
        Procedure of the baseline offloading-based batched inference~\cite{flexgen,deepspeedinf} for the (a) prefill and (b) decoding stages. 
    }
    \label{fig:background}
\end{figure}

\textbf{Prefill Procedure.}
In the prefill stage (\cref{fig:background}(a)), \circled{A} the GPU loads the attention layer weights. 
\circled{B} 
the inputs are multiplied by these weights to generate the query, key, and value matrices (\texttt{QKV}), optionally followed by a positional embedding~\cite{rope}. 
\circled{C} These matrices are used to calculate attention scores for the input prompt sequence via multi-head self-attention computation (\texttt{ATN}).
\circled{D} Next, the MLP weights are loaded from host memory.
\circled{E} The final output activations of the transformer block are generated (\texttt{MLP}).
\circled{F}~The generated KV cache is written back to the host memory or to SSD. 
Steps \circled{A}--\circled{F} are repeated for each transformer block. 

\textbf{Decoding Procedure.}
In the decoding stage (\cref{fig:background}(b)), \circled{1} the GPU retrieves the attention layer weights and the output of the previous layer.
\circled{2} The GPU performs the QKV projection.
\circled{3} The stored KV cache for each layer is loaded. 
\circled{4} The self-attention (\texttt{ATN}) is performed using the loaded KV cache.
\circled{5} The weights ($W_O$, $W_{MLP}$) for the MLP computation are loaded onto the GPU.
\circled{6} The GPU processes the MLP computation.
\circled{7} The KV cache for the newly generated token is written back to its original location for future iterations.
This sequence of steps (\circled{1}--\circled{7}) is repeated for each transformer block until sequence completion. 

\subsection{Near-Storage Processing Devices}
\label{sec:background3}

Computational storage devices (CSDs)~\cite{active_disk, idisk, smartsage, omnicache, NSC-FPGA, nearpm, genstore, instattn} have been extensively studied over the past decades.
By co-locating compute units with storage, CSDs reduce data transfer latency and alleviate I/O bandwidth bottlenecks.
Building on this line of research, commercial implementations~\cite{netezza, exadata, smartssd} have converged on a near-storage processing (NSP) architecture, which typically features a lightweight accelerator (e.g., an FPGA) connected to the storage media via a dedicated internal PCIe interconnect.

A state-of-the-art example is Samsung's SmartSSD~\cite{smartssd}, which features on-board DRAM that serves as a cache for the NAND flash and is accessible by both the host CPU and the device's integrated FPGA. 
The host can perform standard I/O to the SSD or offload computational kernels (e.g., device binary file) to the FPGA via OpenCL APIs~\cite{xilinx_opencl_buf}. 
It leverages an internal P2P PCIe path for direct data movement between the NAND flash and on-board DRAM~\cite{nvmmu, spin}.

\section{Motivation}
\label{sec:motiv}

\begin{figure}[t]
    \centering
    \Description{Figure} 
    \includegraphics[width=\columnwidth]{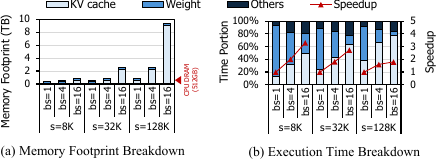}       
    \caption{ 
    Motivational experiments with a 175B model.
    } 
    \label{fig:motiv1}
\end{figure}

This section identifies KV cache I/O as the primary bottleneck in \baseline and motivates our NSP-based solution.
Using the half-precision OPT-175B~\cite{opt}, we analyze performance across varying context lengths ($s$) and batch sizes ($bs$); see \cref{sec:environment} for setup details.


\subsection{Performance Bottleneck Analysis}
\label{sec:motiv1}
Many existing systems~\cite{flexgen, infinigen, deepspeedinf} employ large-batch offline inference to mitigate the frequent model weight transfer overhead.
This approach, however, suffers from the KV cache growing proportionally to both batch size and context length. 

We demonstrate this bottleneck on a state-of-the-art offline inference system~\cite{flexgen} on an A100 server (512GB host memory, four PCIe 4.0 SSDs).
The results in \cref{fig:motiv1}(a) show that the KV cache dominates the system's memory footprint, reaching a terabyte scale that exceeds the typical host memory capacity. 
More critically, \cref{fig:motiv1}(b) reveals that for long-context inference, the system becomes overwhelmingly bottlenecked by data movement.
Transferring the KV cache consumes over 60\% of the total execution time. 
Consequently, this diminishes the advantage of the conventional batching strategy, as there is less room for batching to improve from a reduced portion of weight transfer. 



\subsection{Advantages of Near-Storage Processing}

\begin{figure}[t]
    \centering
    \Description{Figure} 
    \includegraphics[width=\columnwidth]{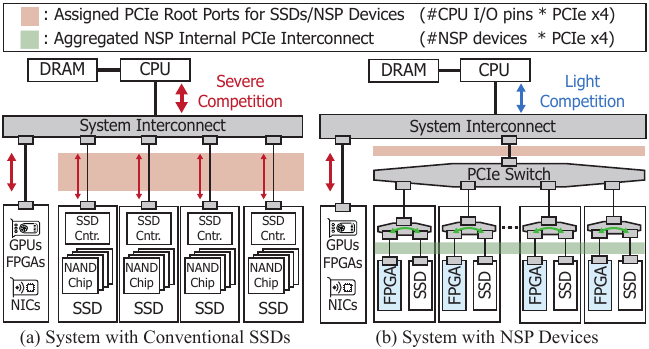}       
    \caption{ 
    An example of a PCIe topology with (a) conventional SSDs and (b) near-storage processing devices.
    } 
    \label{fig:motiv2}
\end{figure}

The study above reveals that the KV cache transfer overhead is a fundamental bottleneck for \baseline. 
This motivates exploration of an NSP solution to address this challenge through several key advantages.

First, the large capacity of NSP devices 
accommodates the terabyte-scale KV caches. 
Second, NSP mitigates the host PCIe bandwidth bottleneck inherent in conventional multi-SSD systems (\cref{fig:motiv2}(a)).
By processing data near storage, NSP minimizes host-device data transfers (\cref{fig:motiv2}(b)). 
%
Third, NSP alleviates contention for host resources already heavily utilized by weight loading and self-attention computation.

Although NSP appears promising, 
a naive deployment does not guarantee performance improvements.
Realizing its full potential requires addressing several key challenges: (1) determining which computations to offload (\cref{sec:ans}), (2) efficiently scheduling tasks between the host and NSP devices (\cref{sec:xcache}), (3) effectively hiding storage write latency (\cref{sec:wb}), and (4) designing a lightweight yet high-throughput storage-side accelerator (\cref{sec:arch}). 
We address these challenges with a complete system built from off-the-shelf components~\cite{smartssd, falcon}.


\section{\thiswork Design}

\subsection{\ANS}
\label{sec:ans}

As discussed in \cref{sec:motiv}, the \baseline suffers from substantial KV cache communication overhead during decoding.
With conventional storage devices, this communication must traverse the shared system interconnect. 
\thiswork addresses this issue with \emph{\ans} (ANS), which offloads attention to a custom accelerator near storage.
It confines the high-volume KV cache traffic to the device's internal path, allowing only the attention inputs and outputs to traverse the shared system interconnect.

\begin{figure}[t]
    \centering
    \Description{Figure.}        
    \includegraphics[width=\columnwidth]{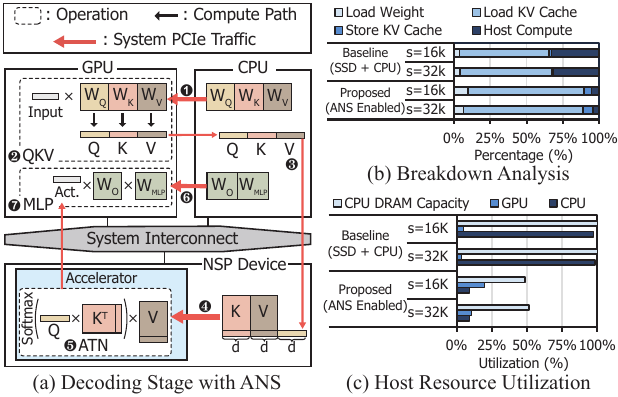}       
    \caption{
    (a) Proposed decoding workflow with an NSP device.
    (b) Latency breakdown of the decoding stage.
    (c)~Host resource (CPU, GPU, DRAM) utilization comparison.
    }
    \label{fig:ans}
\end{figure}

As illustrated in \cref{fig:ans}(a), ANS targets the decoding stage of the inference. 
\bcircled{1} The GPU loads the required LLM weights from the host memory and \bcircled{2} performs the QKV projection.
\bcircled{3} The resulting query, key, and value vectors are written to the SSD. 
\bcircled{4} The near-storage accelerators load the full KV cache from local storage to their DRAM via a private path, \bcircled{5} perform the attention computation, and transfer the final attention output back to the host.
\bcircled{6} The GPU then loads MLP weights from the host memory and \bcircled{7} executes the subsequent MLP layers.

This strategy yields a significant reduction in data movement.
With a context length $s$ and hidden dimension $h$ using half precision, the baseline's interconnect read traffic for the KV cache is $4\cdot s \cdot h$ bytes per decoding step.
\Ans eliminates this traffic, replacing it with a transfer of only the $2\cdot h$-byte attention output vector from the device to the host.
This benefit comes at the cost of a minor increase in write traffic, from $4\cdot h$ bytes in the baseline (new K and V) to $6\cdot h$ bytes in \ans (new Q, K, and V).

The overall traffic reduction is substantial because the dominant read traffic scales with the context length $s$.
Comparing the baseline traffic $T_{BASE}$, and the traffic with \ans $T_{ANS}$,
the benefit grows linearly with $s$:
\begin{equation}
\begin{aligned}
    \frac{T_{BASE}}{T_{ANS}} = \frac{4sh + 4h}{2h + 6h} 
    =\frac{s+1}{2} > 1\,(\because s > 1).
\end{aligned}
\end{equation}

To scale this approach to multiple NSP devices, we parallelize attention along both the batch and attention head dimensions.
With batched inference, the product of the batch size and the number of attention heads provides sufficient dimensions to distribute the workload across the available NSP devices.
For this, the prefill stage distributes the KV cache in the same parallelization across the NSP devices. 

Our approach follows the principle of near-storage processing~\cite{smartssd,active_disk,idisk,smartsage,omnicache} and effectively reduces interconnect traffic.
However, in practice, the offloading of attention operations to NSP shifts the performance bottleneck, now causing the internal storage I/O to dominate end-to-end latency, as shown in \cref{fig:ans}(b).
This poses a key challenge for practical NSP systems: reducing internal storage I/O.
Fortunately, the offloading leaves the host severely underutilized (<20\%, \cref{fig:ans}(c)), with the host remaining active only during the computationally inexpensive QKV projection and MLP stages.
Based on this observation, the following section proposes a strategy to ensure effective cooperative execution between the host and the near-storage device, addressing the aforementioned challenge.

\subsection{Cooperative X-Cache}
\label{sec:xcache}

\begin{figure}[t]
    \centering
     \Description{Figure.} 
    \includegraphics[width=\columnwidth]{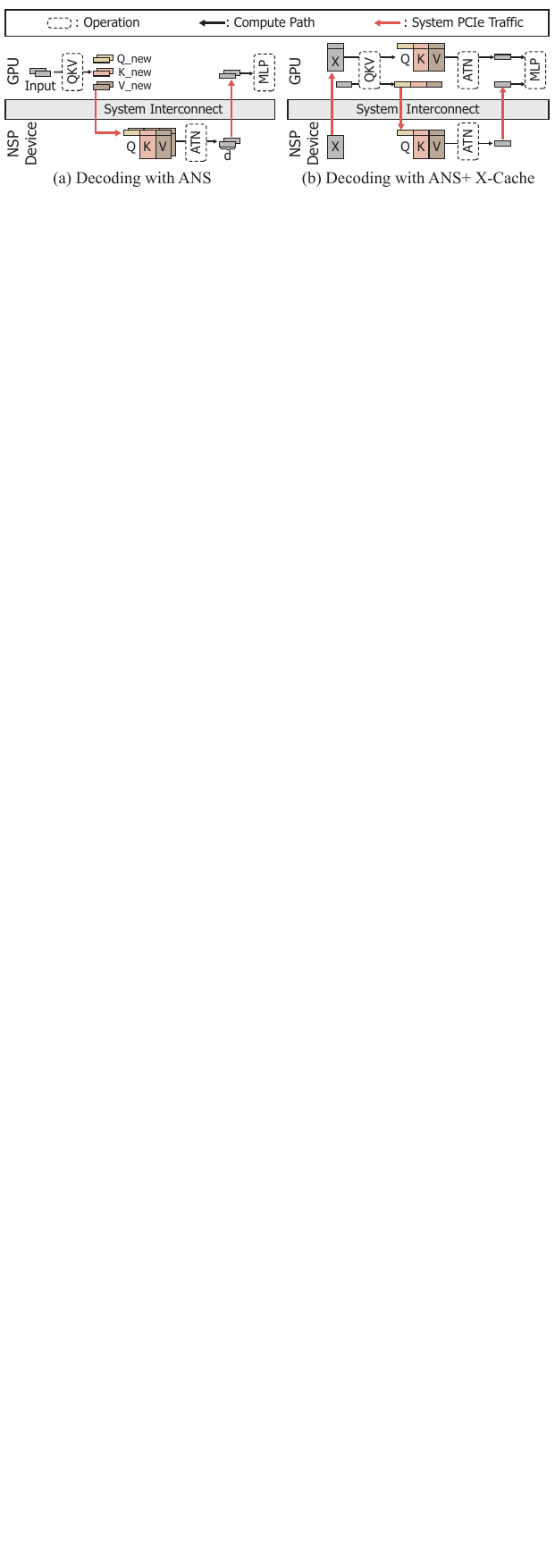}       
    \caption{
   Decoding procedure with (a) only ANS and (b) both ANS and cooperative \xcache.
    }
    \label{fig:xcache}
\end{figure}

To reduce the internal storage I/O by exploiting the host underutilization of the baseline procedure (\cref{fig:xcache}(a)), we propose a cooperative execution schedule that partitions the attention workload between the host and the NSP devices.
Recalling from \cref{eq:qkv} that the key and value caches are projections of the input activation $X$, 
we cache the pre-projection input activation $X$ for the historical context instead of the derived K and V tensors.
We term this technique \emph{cooperative \xcache}.

\cref{fig:xcache}(b) details the procedure.
We define $\alpha$ as the fraction of batches managed by \xcache.
The remaining $1-\alpha$ portion utilizes the standard KV cache stored in the SSD.
This parameter $\alpha$ partitions the batch ($b$) and head ($h$) dimensions rather than the sequence dimension ($s$).
During the prefill phase, the system persists the input activations $X$ corresponding to the $\alpha$ portion.
In the decoding phase, the GPU loads these \xcache entries directly via GPU-Direct Storage (GDS)~\cite{gds} to minimize system traffic.
The GPU subsequently regenerates the required $K$ and $V$ tensors by re-executing the projection operations (\cref{eq:qkv}).
Simultaneously, the near-storage accelerator (\cref{sec:ans}) performs multi-head attention on the $1-\alpha$ portion of the KV cache.

Since \xcache ($b\times~h\times~s\times~d$) requires only half the storage capacity of the combined key and value tensors ($2\times~b\times~h\times~s\times~d$), it effectively halves both the system-interconnect and flash-read traffic.
This presents a favorable trade-off, as it utilizes the GPU's high processing power that would otherwise remain idle while attention operations are offloaded.
Since this recomputation is executed concurrently with the attention operation on the NSP devices, its latency is effectively hidden.
In addition, this positively affects the SSD endurance from less writes to the SSDs (\cref{exp:endurance}).

\textbf{I/O Analysis.}
To quantify the impact of cooperative \xcache, we derive a first-order cost model.
Assuming 16-bit precision (2 bytes per element), 
the time to transfer the \xcache portion ($\alpha$) from the NSP device to the GPU over the host interconnect with bandwidth $B_{PCI}$ is
\[
T_{\text{PCI}} = \nicefrac{ \alpha \cdot s \cdot h \cdot 2 }{B_{\text{PCI}}}.
\]

Let $C_{\text{GPU}}$ denote the GPU’s compute capability in FLOPS.  
To regenerate $K$ and $V$, the input $X \in \mathbb{R}^{s \times h}$ is multiplied by the weight matrices $W_K, W_V \in \mathbb{R}^{h \times h}$.  
The time required for this additional recomputation per transformer block is
\[
T_{\text{GPU}} = \nicefrac{ \alpha \cdot s \cdot h^{2} \cdot 2 }{C_{\text{GPU}}}.
\]

The storage read time is governed by the SSD's bandwidth, $B_{SSD}$, as a sum of reading the \xcache data ($\alpha$ fraction) and the KV-cache data ($1-\alpha$ fraction):
\[
T_{\text{SSD}} = \nicefrac{(\alpha \cdot (s \cdot h \cdot 2) + (1-\alpha) \cdot (2 \cdot s \cdot h \cdot 2))}{B_{\text{SSD}}}.
\]

Assuming that the regeneration computation and data transfers are well-pipelined and can be overlapped,
\[
T_{\text{effective}} = \max~\!\bigl(T_{\text{GPU}},\, T_{\text{SSD}},\, T_{\text{PCI}}\bigr).
\]

Because $T_{\text{GPU}}$ is often small, 
an $\alpha$ satisfying $T_{\text{SSD}} = T_{\text{PCI}}$ typically yields the best latency.
Let $S_X$ denote the size of $X$, which implies the KV-cache size is $S_{KV} = 2\cdot~S_X$.
Setting $T_{\text{PCI}} = T_{\text{SSD}}$ yields
$$
\frac{\alpha S_X}{B_{\text{PCI}}} = \frac{\alpha\cdot~S_X + (1-\alpha)\cdot~S_{KV}}{B_{\text{SSD}}} = \frac{\alpha\cdot~S_X + 2\cdot~(1-\alpha)\cdot~S_X}{B_{\text{SSD}}}.
$$

Canceling $S_X$ and solving for $\alpha$ yields
$$
\alpha = \frac{2 B_{\text{PCI}}}{B_{\text{SSD}} + B_{\text{PCI}}}.
$$
This implies that the ratio between $B_{\text{SSD}}$ and $B_{\text{PCI}}$ determines the optimal $\alpha$.  
We model $B_{SSD}$ to scale linearly with the number of NSP devices,
while the effective bandwidth $B_{PCI}$ is profiled over several layers.
We use this formulation to automatically select an $\alpha$ 
closest to a power of two. 

\subsection{\WB}
\label{sec:wb}
\begin{figure}[t]
    \centering
         \Description{Figure.} 

        \includegraphics[width=\columnwidth]{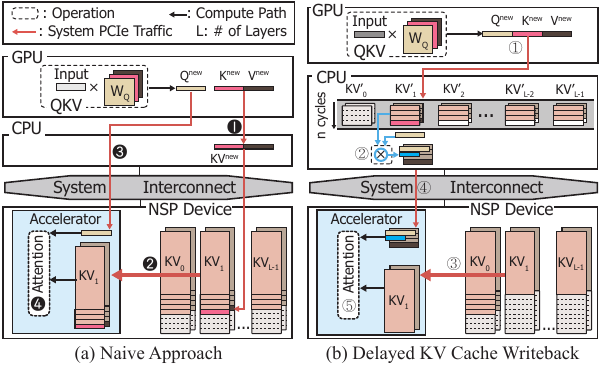}     
    \caption{
     KV-cache write-back procedure with (a) naive CSD offloading and (b) \wb.
    }
    \label{fig:wb}
\end{figure}

While \ans and cooperative \xcache improve overall throughput, handling write-backs during decoding phase remains challenging.
During the prefill phase, KV caches follow a \emph{row-wise layout} ($b\times~h\times~s\times~d$).
The X-cache also employs the same mechanism, sharing an identical size and layout with the K-cache.
This layout ensures high SSD bandwidth because the minimum access granularity ($s\times d$) typically exceeds 4KiB.
Conversely, each decoding iteration generates small KV vectors ($1\times~d$) that must be appended to the existing cache ($s\times~d$) on storage.
For these small vectors to be truly available for subsequent attention operations, they must be committed to storage. 
This dependency places storage write latency on the critical execution path and incurs an inefficient sub-page granularity writes.


In the naive approach shown in \cref{fig:wb}(a), after the GPU generates new KV entries, \bcircled{1} each entry is individually written to the SSD via direct I/O.
\bcircled{2} Subsequently, the processing unit reads the entire updated KV cache from storage to its local memory.
\bcircled{3} Once the new query vector is transferred to the processing unit, \bcircled{4} the attention operation is performed.
In this method, SSD writes lie on the critical execution path.
Moreover, each KV entry (256 bytes) is far smaller than the SSD page size (4KiB), leading to poor write performance. 

To address this, we propose \emph{\wb}, shown in \cref{fig:wb}(b).
Instead of writing every newly generated KV entry directly to storage, we keep dedicated buffers for KV data in host memory and allow the buffer to feed directly into the accelerator’s off-chip memory. 
From this, we obtain the benefits of
(1) the write latency stays off the critical path, and
(2) data are written in page-sized chunks. 

A remaining issue from the \wb is that the same KV vectors are sent redundantly 
until they are spilled.
To mitigate this overhead, the host CPU precomputes the partial dot products ($QK^T$) between the current query and the buffered key vectors.
Only the resulting scalar values (marked blue) are sent to the processing unit. 

The optimized workflow proceeds as follows:
\circled{1} New KV entries are staged in host memory buffers.
\circled{2} The CPU precomputes the partial $QK^T$ scores from these buffered entries.
\circled{3} The already stored K and V tensors are read from the SSD into the accelerator.
\circled{4} The query, the precomputed $QK^T$ scalars, and the new V entries are transferred from host memory to the accelerator.
\circled{5} The accelerator uses these components to complete the final attention computation.
The buffered values are later spilled to the storage after a predetermined \emph{spill interval}. 
Given that the SSD write granularity is 4KiB and each KV entry is typically 256 bytes per head~\cite{opt, llama, llama2, llama3, deepseekv3, mixtral, glam}, performance optimization is typically achieved with the spill interval set to 16. 
   
\subsection{Attention Accelerator Architecture}
\label{sec:arch}

\begin{figure}[t]
     \Description{Figure.} 

    \centering
    \includegraphics[width=\columnwidth]{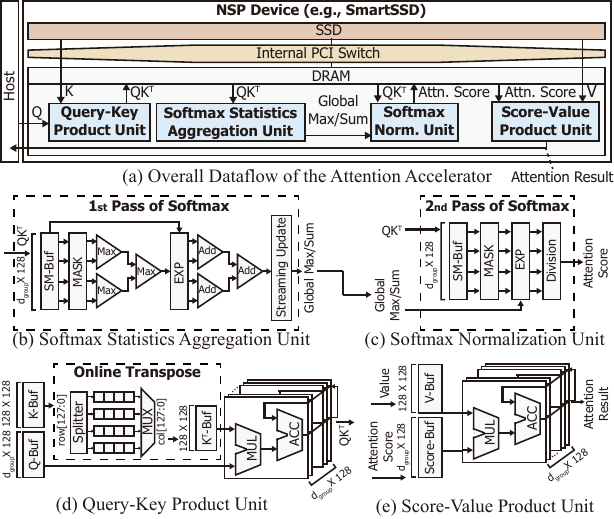}
    \caption{Dataflow and units of the attention accelerator.}
    \label{fig:arch}
\end{figure}

While there are several existing transformer accelerators~\cite{splitwise,neupims,attacc,duplex, spatial, dfx}, \thiswork designs a new architecture to address the following challenges for near-storage processing:
\begin{enumerate}[wide=0pt, labelindent=0pt, label=\arabic*.]
    \item \emph{Limited available resources.}
    Near-storage accelerators have tight resource budgets. 
    For example, Samsung SmartSSD employs a moderately sized FPGA, 
    which is further limited by the SSD power budget~\cite{pci}.
    This becomes especially problematic for softmax operations for long contexts.
    \item \emph{Layout conflict for key writes and reads.}
    As described in \cref{sec:wb}, the new KV entries are added to the existing KV cache in a row-wise manner.
    However, the key matrix has to be read in columns (i.e., $K^T$ for the attention), creating a mismatch.
    \item \emph{Support for attention variants.} 
    Modern LLM architectures are increasingly adopting variants of attention such as group-query attention (GQA).
    While it can be processed with existing architectures for multi-head attention, it would cause redundant data accesses.
\end{enumerate}

\begin{algorithm}[t]
    \small
    \caption{Efficient Softmax Implementation}
    \DontPrintSemicolon
    \SetNoFillComment

    $m \leftarrow -\infty$, $Z \leftarrow 0$; \tcp*[f]{\small Initialize running max and sum}

    \tcp{\small First pass: compute global statistics}
    \ForEach{block $B$ in input sequence $x$}{
        $B \leftarrow \text{MASK}(B)$;~~~$m_B \leftarrow \max_{b \in B} b$;\;
        $S_B \leftarrow \sum_{b \in B} \exp(b - m_B)$;\;       
        \tcp{\small Streaming Update Unit }
        \eIf{$m_B > m$}{
            $Z \leftarrow Z \cdot \exp(m - m_B) + S_B$; \;
            $m \leftarrow m_B$; \;
        }{
            $Z \leftarrow Z + S_B \cdot \exp(m_B - m)$; \;
        }
    }

    \tcp{\small Second pass: normalize element by element}
    \ForEach{$x_i$ in input sequence $x$}
    {
        $x_i \leftarrow \text{MASK}(x_i)$;~~~$y_i \leftarrow \exp(x_i - m) / Z$; \;
    }

\label{algo:softmax}
\end{algorithm}

\cref{fig:arch}(a) depicts the overall dataflow of the proposed accelerator.
Existing spatial architectures~\cite{spatial2, spatial3, spatial4} compute the entire attention operation in a single pass, requiring on-chip memory capacity that grows prohibitively large with sequence length~\cite{instattn, spatial}.
In contrast, we adopt a temporal architecture~\cite{dfx, npe, ftrans} that processes the attention computation in blocks (e.g., 128 tokens per block).
As illustrated in \cref{fig:arch}(a), four hardware units form a concurrent pipeline to process each block in accordance with data dependencies.
This choice substantially reduces the resource requirement, especially the on-chip memory. 

\textbf{Two-Pass Softmax.}
Given an input vector \(x = [x_1,\dots,x_s]\), the softmax operation is
\begin{equation}
\begin{aligned}
\text{softmax}(x) &= [o_1,\dots,o_s], \\
\text{where } o_i &= \frac{e^{x_i-\max(x)}}{\sum_{j=1}^{s} e^{x_j-\max(x)}} \quad \text{for } i=1,\dots,s.
\end{aligned}
\label{eq:softmax}
\end{equation}

To avoid numerical instability, this is typically done in three passes~\cite{online}: one to find the global maximum $max(x)$, a second to compute the sum of exponentials, and a third for element-wise normalization.
For long sequences, 
this method creates a major bottleneck from the off-chip memory traffic. 

\begin{sloppypar}
To address this, inspired by the online softmax algorithm~\cite{flash_attn1, flash_attn2, online}, we design a two-pass (instead of three) method as in \cref{algo:softmax}.
The two-pass softmax is conducted via two units: softmax statistics aggregation unit (\cref{fig:arch}(b)) and softmax normalization unit (\cref{fig:arch}(c)).

In the first pass (\cref{fig:arch}(b)), the statistics aggregation unit processes the input vector in blocks of $d_{group} \times 128$ elements.
Each block is streamed through a pipelined max reduction tree to compute a local maximum (line 3).
The key enabler of this two-pass algorithm is that
this local maximum is used instead of the global maximum to stabilize the exponential values by feeding it into parallel exponentiation units. 
The results are aggregated by the adder tree to compute a partial sum (line 4).
Then, depending on whether the global maximum is updated, the \texttt{streaming update} unit modifies the global sum, which serves as the final denominator (lines 5-9). 
The second pass (\cref{fig:arch}(c)) performs the normalization, composed of element-wise exponential and division (lines 10-11).
For the masking of the attention calculation, \texttt{MASK}s in both units apply padding masks and precomputed scalars from the host (\cref{sec:wb}) (lines 3, 11) for each input element.
\end{sloppypar}
 

\textbf{GEMV with Online Transpose.}
The GEMV units perform the query-key (\cref{fig:arch}(d)) and attention score-value (\cref{fig:arch}(e)) products.
A central challenge in the query-key computation is the memory layout transformation for the transposed Key matrix ($K^T$).
Naive solutions would be to store in a transposed format $K^T$ or in both. However, they would be inefficient for SSD writes or add redundant writes.

Leveraging the block-based mechanism, our design integrates an efficient in-place transposition module, as shown in \cref{fig:arch}(d). 
By simply performing a local block transpose and matching it with the right $Q$ block, global transpose can be avoided.
Thus, the unit loads a $128 \times 128$ square block of the Key matrix into an on-chip buffer (\texttt{K-Buf}), performs the transposition locally, then stores the transposed matrix to another on-chip buffer (\texttt{K$^T$-Buf}).
Finally, the transposed matrix is streamed to the MAC units for a blocked GEMV.

\textbf{Native Support for Attention Variants.}
Our architecture also provides native support for attention variants such as GQA~\cite{llama2,llama3,qwen2}, where $d_{group}$ query heads share a single KV cache.
For this, the K/V buffers broadcast their values to the $d_{group} \times 128$ parallel MAC units to concurrently process $d_{group}$ query vectors that share access to the same KV cache data.
Without this feature, the shared KV Cache would be read redundantly for each query in a group.
The partial results are accumulated into each final output buffer per query, reducing redundant off-chip memory traffic.

\section{Full System Integration}
\label{sec:impl}

\begin{figure}[t]
    \centering
     \Description{Figure.} 

    \includegraphics[width=\columnwidth]{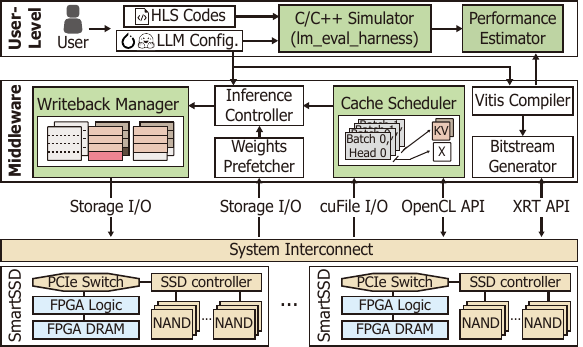}       
    \caption{
    Overview of \thiswork, a ready-to-use framework. 
    }
    \label{fig:overview}
\end{figure}

We prototype \thiswork on a real system using commercially available SmartSSDs and a PyTorch-based framework. 
\cref{fig:overview} illustrates the overview of \thiswork, which comprises three main components: a user-level design flow for accelerator customization (top), a middleware layer for \thiswork (middle), and the underlying hardware testbed (bottom).

\subsection{User-Level Design Flow}
\label{sec:host_impl}

\thiswork provides an end-to-end design flow for generating FPGA bitstreams for the SmartSSDs, as depicted in the top of \cref{fig:overview}.
While we provide pre-compiled binaries for common use cases, users can customize the high-level synthesis~\cite{vitis_hls, intro_vlsi} (HLS) source code to implement specialized attention variants~\cite{deepseekv3, gemini15}.
For easy customization, we offer a simulation and verification tool integrated with the \texttt{lm-evaluation-harness}~\cite{eval_harness}.
This allows users to validate the functional correctness of accelerator designs on standard LLM benchmarks~\cite{copa,openbookqa,wikitext2, longbench} before committing to resource-intensive synthesis.
Furthermore, we provide a performance estimator for the accelerator based on the cycle counts and the clock frequency obtained from HLS. 
Across sequence lengths from 4K to 32K, the estimator achieves a Pearson correlation coefficient of 0.93 compared to measured hardware throughput for the three kernels (\cref{tab:imple_result}).
This high correlation demonstrates that the simulator accurately captures the performance trends of the target hardware.

\subsection{Middleware Layer}
The middleware layer (\cref{fig:overview}) orchestrates the LLM inference pipeline.
During runtime, the \emph{Inference Controller} directs core LLM operations (e.g., MLP, QKV) while the \emph{Weights Prefetcher} stages model weights from the host to the GPU.
The \emph{Cache Scheduler} (\cref{sec:xcache}) determines an optimal \xcache ratio $\alpha$, partitions the KV cache between SmartSSDs and the GPU during prefill, and coordinates parallel execution during decoding.
Concurrently, the \emph{Writeback Manager} (\cref{sec:wb}) asynchronously spills data to the SmartSSDs upon reaching a configured threshold, preventing LLM inference pipeline stalls.
These C++ runtime components interface with PyTorch via pybind11~\cite{pybind11} and are packaged as a callable Python module using distutils~\cite{distutils}.
An offline toolchain, featuring the \emph{Vitis Compiler} and \emph{Bitstream Generator}, synthesizes custom HLS designs into FPGA bitstreams for deployment.

\subsection{Hardware Testbed}
The \thiswork testbed consists of up to 16 SmartSSDs, each equipped with our custom accelerator (\cref{sec:arch}).
These devices are integrated into a host system via a general-purpose PCIe expansion chassis~\cite{falcon} that supports 16 PCIe 4.0 lanes.
To enable peer-to-peer (P2P) communication~\cite{nvmmu,spin}, we allocate buffers in the FPGA's on-board DRAM using the Xilinx OpenCL extension \texttt{CL\_MEM\_EXT\_PTR\_XILINX}.
This flag makes the buffer memory-mapped and accessible to the CPU, FPGA, and the SSD controller, facilitating direct data transfers.
Each FPGA connects to its corresponding SSD via an internal PCIe switch, ensuring that all communication remains within the SmartSSD and avoids inter-device traffic.

\subsection{FPGA Implementation Details}
\label{sec:fpga_details}

This section details the FPGA logic implementation and the specific design decisions for hardware bottlenecks.

\textbf{Computation Design and Hardware Trade-offs}
We utilize the AMD Vitis HLS Math Library for hardware optimized \texttt{cmath} operations, including exponential functions and data type conversions.
Conversely, we develop the softmax and GEMV modules from scratch to enable fine-grained architectural control within the resource constraints of a near-storage accelerator.
A primary design decision involves selecting unrolling factors that maximize throughput without exceeding hardware limits.
Specifically, we apply a loop unrolling factor of two to the exponential units.
This configuration balances the high DSP consumption of exponential operations with the necessity for parallel execution.
Furthermore, we implement maximum value and sum-of-exponential reductions using a two-level tree structure with a depth of four.
This design choice minimizes pipeline latency while ensuring timing closure for the high-latency four-way reduction logic.
The GEMV operations are parallelized across 128 MAC units.
While the hardware resources permit higher degrees of parallelism, we utilize 128 units because this configuration successfully saturates the available DRAM bandwidth.

\textbf{Memory Access Optimization}
To maximize hardware interface efficiency, input sequences are zero-padded to multiples of 32 to facilitate AXI burst transactions.
We utilize the HLS Vector Library to implement 32 element bursts of 16 bit half precision data to align each transaction with the 512 bit AXI4 data width.
This configuration leverages the maximum bitwidth supported by the hardware to facilitate high throughput memory access.
We also apply cyclic partitioning with a factor of 32 to the BRAM buffers for DRAM reads to match the 512-bit granularity.
These techniques collectively ensure that the DRAM bandwidth is fully utilized.

\textbf{Dataflow Integration}
We apply the \texttt{DATAFLOW} pragma to the top-level kernel to enable task-level pipelining.
This allows the system to overlap KV cache loading with the computation of preceding segments.
To meet strict dataflow constraints, our design follows a forward-only data propagation strategy.
For example, although the softmax input/output buffers and the KV cache buffers share identical dimensions, we allocate dedicated hardware resources for each module rather than employing shared buffers.
This design choice eliminates memory access contention and satisfies the requirements of the \texttt{DATAFLOW} pragma, enabling the HLS compiler to achieve effective task-level parallelism.

\textbf{Numerical Precision and Stability.}
The design utilizes native FP16 data types for storage while performing intermediate calculations in FP32.
Specifically, accumulations and exponential operations use FP32 to preserve numerical stability.
During the attention mechanism, a masking module assigns a constant value of $-10^4$ to padding tokens.
This ensures that these tokens do not influence softmax results, thereby maintaining numerical stability during inference.

\section{Evaluation}

\subsection{Experimental Setup}
\label{sec:environment}
\begin{table}
\footnotesize
\centering
\caption{Experimental Environments}
\label{tab:setup}
 \def\arraystretch{1.0}
\resizebox{\columnwidth}{!}
{
\setlength{\tabcolsep}{15pt}
\begin{tabular}{ccc}
\toprule
\multirowcell{8}[-0.4ex]{\textbf {HW}} 
&GPU & NVIDIA A100 (40GB), H100 (80GB)  \\
\cmidrule(lr){2-3} 
&CPU & Xeon(R) Gold 6342, 24C 48T  \\
\cmidrule(lr){2-3} 
&Memory& 16$\times$32GB DDR4-3200\\
\cmidrule(lr){2-3} 
&CSD & SAMSUNG SmartSSD~\cite{smartssd}, 3.84TB   \\
\cmidrule(lr){2-3} 
&Conventional SSD & SAMSUNG PM9A3~\cite{pm9a3}, 3.84TB \\
\cmidrule(lr){2-3} 
&PCIe Expansion & H3 Falcon 4109  \\

\midrule

\multirowcell{5}[-0.8ex]{\textbf {SW}} 
& OS / Python & Ubuntu 20.04 LTS / 3.12.4 \\ 
\cmidrule(lr){2-3} 
& PyTorch / DeepSpeed    & 2.4.1 / 1.15.1 \\ 
\cmidrule(lr){2-3} 
& Vitis / XRT & 2023.1 / 2.12.427 \\
\cmidrule(lr){2-3} 
& CUDA / OpenCL  & 12.1.105 / 2.2 \\ 

 \bottomrule
\end{tabular}
}
\end{table} 
\begin{table}[t]
  \centering
  \caption{\label{tab:model} Model Configurations}
  \resizebox{\columnwidth}{!}{%
      \setlength{\tabcolsep}{2pt}
  \begin{tabular}{c|ccccccc}
    \toprule
    Model & \# layers & Hidden & Interm. &\# heads & \# KV heads & $d_{\text{group}}$ &  \# experts \\
    
    \midrule
          OPT-30B    &  48  & 7,168 & 28,672 & 64 & 64 (MHA)  & 1  & - \\
          OPT-66B    &  64  & 9,216  & 36,864 & 72 & 72 (MHA) & 1  & - \\
          OPT-175B   &  96  & 12,288 & 49,152 & 96 & 96 (MHA) & 1  & - \\
    \midrule
          Qwen2.5-32B &  64  & 5,120 & 27,648 & 40 & 8 (GQA) & 5    & - \\
          Mixtral-8x7B & 32 & 4,096 & 14,336 & 32 & 8  (GQA) & 4   &  8  \\
          GLaM-143B  & 32 & 4,096 & 16,384 & 32 & 32 (MHA)  & 1  & 64 \\
    \bottomrule
    \addlinespace
  \end{tabular}%
  }
\end{table}

\textbf{Hardware Configuration.} 
Our experimental setup is detailed in \cref{tab:setup}. 
Each SmartSSD~\cite{smartssd} includes a 3.84TB NVMe SSD and a Kintex UltraScale+ KU15P FPGA~\cite{kintex} 
attached with 4GB of DDR4-2400 DRAM. 
Our setup also includes either an A100 or H100 GPU, 
connected to the CPU via PCIe 4.0$\times$16 lanes.
The baseline systems are equipped with four 3.84TB SAMSUNG PM9A3 SSDs~\cite{pm9a3}, each providing up to 6,900MB/s read and 4,100MB/s write bandwidth.
Each SSD is connected to the host via a dedicated PCIe 4.0 x4 link, occupying a total of 16 PCIe lanes for storage—matching the total number of host PCIe lanes used by the SmartSSDs.

\textbf{Inference System Configurations.}
We compare \thiswork against two state-of-the-art \baseline frameworks, FlexGen~\cite{flexgen} and DeepSpeed~\cite{mii,deepspeedinf}.

We evaluate three FlexGen configurations for KV cache offloading: to host DRAM (\textbf{\texttt{FLEX(DRAM)}}), to four PCIe 4.0 SSDs (\textbf{\texttt{FLEX(SSD)}}), and to the 16 NVMe SSDs in our SmartSSD platform with the FPGAs disabled (\textbf{\texttt{FLEX(16 PCIe 3.0 SSDs)}}).
For DeepSpeed, we use its ZeRO-Inference engine~\cite{deepspeedinf} and extend it with Unified Virtual Memory~\cite{uvm} (\textbf{\texttt{DS+UVM(DRAM)}}).
This extension was necessary to manage the high GPU memory pressure of intermediate activations for long-context inference, as this functionality is not natively supported.
Our system, \textbf{\texttt{\thiswork~(N SmartSSDs)}}, is configured with N SmartSSDs (N=8 by default), an \xcache ratio derived from \cref{sec:xcache}, and a storage spill interval of 16. 
To ensure fair comparisons, all systems utilize FlashAttention~\cite{flash_attn1,flash_attn2} for the prefill stage.
All baseline storage configurations use a software RAID-0 using \textit{mdadm}~\cite{mdadm}, and all baselines offload attention computation to the CPU~\cite{flexgen, deepspeedinf} during decoding.

\textbf{Workload.}  
We evaluate representative decoder-only models listed in \cref{tab:model}.
Our evaluation includes models with standard multi-head attention (MHA), such as OPT-30B, 66B, and 175B~\cite{opt}, which we refer to by their parameter counts.
We also evaluate models with GQA~\cite{gqa}, including Qwen2.5-32B~\cite{qwen2}, and Mixture-of-Experts (MoE) models like Mixtral-8$\times$7B~\cite{mixtral} and GLaM-143B~\cite{glam}, with two active experts per layer.
Although some models were pre-trained with shorter context lengths, we benchmark performance 
up to 128K tokens to highlight our system’s long-context potential, anticipating future models with extended contexts~\cite{llama3,gemini15,gpt4}.
We use a default batch size of 16, FP16 precision, and a generated output length of 64 tokens.
Weights reside in CPU memory when capacity permits, whereas models exceeding 100B parameters are offloaded to storage.

\subsection{Implementation Results}
\label{exp:impl}

\begin{table}[t]
    \centering
    \caption{Resource Utilization and Achieved Performance}
    \label{tab:imple_result}
    \setlength{\tabcolsep}{5pt} 
    \resizebox{\columnwidth}{!}{
    \begin{tabular}{c|ccccc|cc}
        \toprule
        Type 
        & LUT 
        & FF 
        & BRAM 
        & URAM 
        & DSP
        & Peak Perf. 
        & Power \\
        \midrule
        $d_{\text{group}}=1$ & 38.76\% & 28.57\% & 51.02\% & 9.38\%  & 10.06\% & 11.9 GFLOPS & 11.25 W \\
        $d_{\text{group}}=4$ & 56.60\% & 39.70\% & 59.30\% & 9.38\%  & 20.27\% & 46.8 GFLOPS & 15.39 W \\
        $d_{\text{group}}=5$ & 67.40\% & 46.15\% & 58.49\% & 9.38\%  & 27.79\% & 56.3 GFLOPS & 16.08 W \\
        \bottomrule
    \end{tabular}
    }
\end{table}

\begin{figure}[t]
    \centering
    \Description{Figure.} 
    \includegraphics[width=\columnwidth]{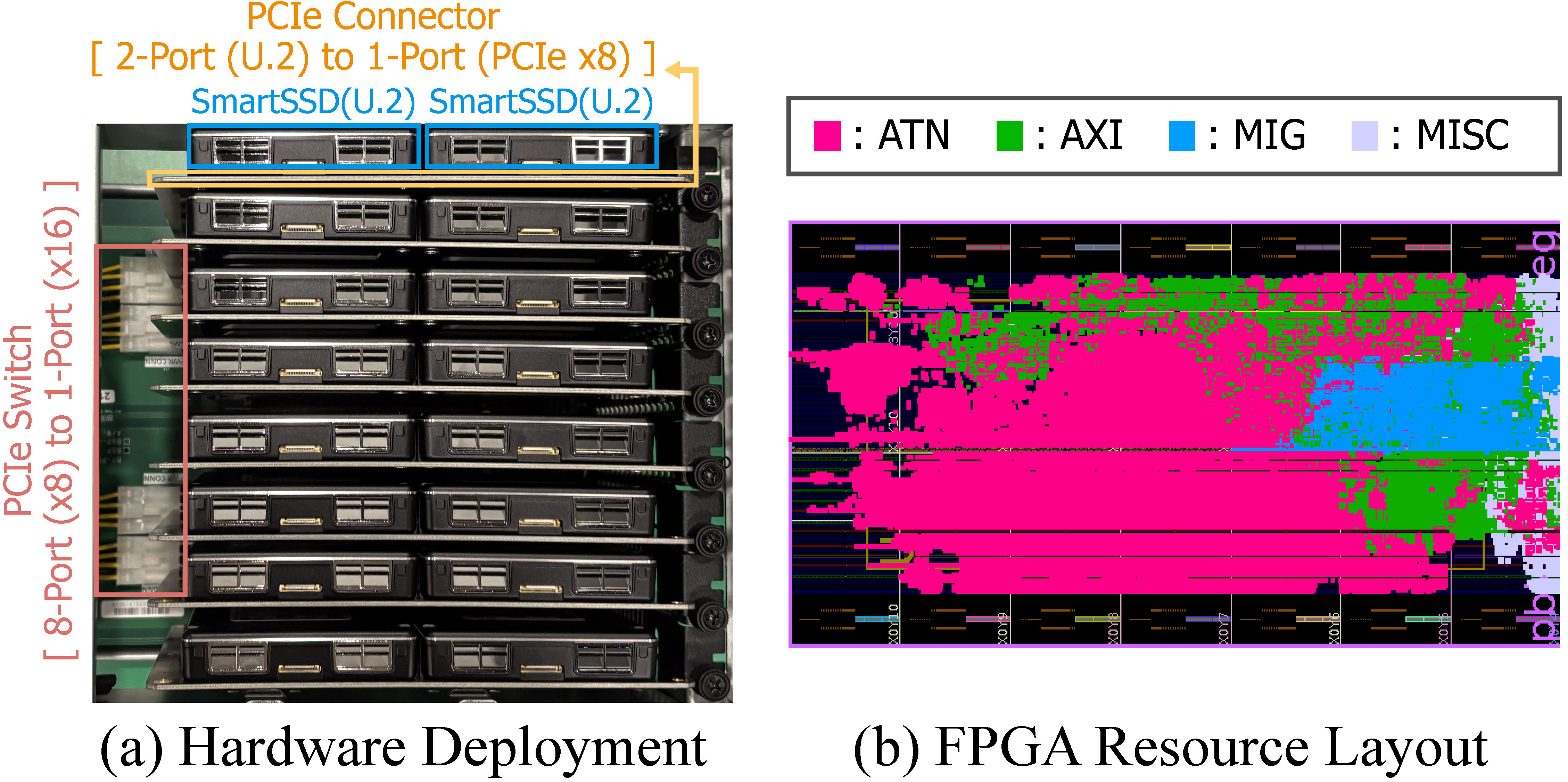}       
    \caption{
    (a) Hardware deployment of the proposed system, featuring Samsung SmartSSDs~\cite{smartssd} interconnected via a PCIe.
    (b) The corresponding FPGA resource floorplan of the user logic partition (ULP) region, for a $d_{group}=1$ configuration. 
}
\label{fig:imple_result}
    
\end{figure}

\cref{fig:imple_result}(a) shows our hardware testbed, which integrates 16 SmartSSDs via a PCIe switch. The switch connects to the host using a PCIe x16 link and provides eight PCIe x8 ports, each driving two SmartSSDs through a U.2 adapter.
\cref{fig:imple_result}(b) presents the FPGA resource floorplan, and \cref{tab:imple_result} details resource utilization, peak performance, and total on-chip power consumption.
The attention module is the primary consumer of resources.
Specifically, the softmax unit utilizes a large fraction of DSP blocks for exponential computations.
In contrast, GEMV units primarily utilize LUTs to manage complex memory transactions such as transposition.

The reported on-chip power, comprising static, dynamic, and PCIe transceiver power, peaks at 16.08 W per accelerator at $d_{group}=5$.
Consequently, a full 16-accelerator deployment consumes approximately 258 W, which is comparable to the power usage of a single mid-range GPU.
Increasing $d_{group}$ improves throughput by sharing the key-value cache across multiple queries, but this comes at the cost of increased resource and power consumption.
The design achieves an operating frequency of 296.05 MHz.
This frequency approaches the 300 MHz limit imposed by the strict power envelope of the SmartSSD platform.

\subsection{Performance Comparison}
\label{exp:main}

\begin{figure}[t]
    \centering
    \Description{Figure.} 

    \includegraphics[width=\columnwidth]{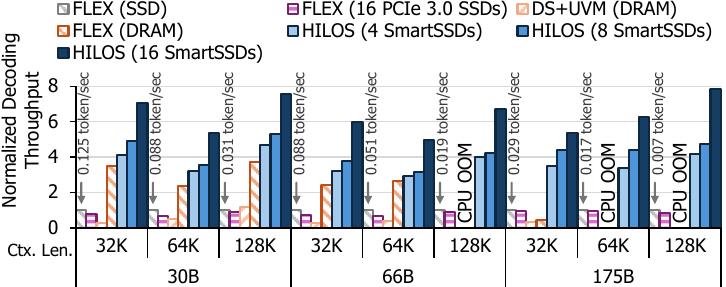}
    \caption{
    Performance comparison of \thiswork and baselines.
    }
    \label{fig:main}
\end{figure}

\cref{fig:main} compares the LLM decoding throughput of \thiswork against four baselines across model sizes and context lengths. 
All results are normalized to \texttt{FLEX(SSD)}. 
Since the \texttt{DS+UVM(DRAM)} offloads activations using UVM~\cite{uvm}, its overhead results in a slowdown of over 4$\times$ relative to \texttt{FLEX(DRAM)}.
The \texttt{FLEX(16 PCIe 3.0 SSDs)} baseline, using 16 SmartSSDs with the FPGAs disabled, motivates our design.
Without near-data compute units, all KV cache data must be transferred to host memory.
This saturates the PCIe bandwidth, similar to the \texttt{FLEX(SSD)} baseline. 
Consequently, this baseline achieves only 0.64$\times$ to 0.94$\times$ the throughput of \texttt{FLEX(SSD)}.

\texttt{FLEX(DRAM)} performs relatively well with the KV cache stored in CPU memory while reducing the batch size to stay within capacity.
However, this method faces scalability issues on larger models and longer contexts. 
\texttt{FLEX(DRAM)} often encounters memory exhaustion, even at batch size 1.

We evaluate \thiswork using four, eight, and 16 SmartSSDs.
With four devices, \thiswork outperforms \texttt{FLEX(DRAM)} by 1.10$\times$ to 1.36$\times$ by managing the KV cache in storage, which enables larger batch sizes and reduces PCIe traffic.
Scaling to 16 SmartSSDs increases the speedup to a range of 1.88$\times$ to 2.49$\times$.
For long contexts where \texttt{FLEX(DRAM)} fails even with a batch size of one, \thiswork achieves a 5.3$\times$ to 7.8$\times$ speedup over the storage-based \texttt{FLEX(SSD)}.


\subsection{ Sensitivity Test Results }
\label{exp:sensi}

\begin{figure}[t]
    \centering
    \Description{Figure.} 

    \includegraphics[width=\columnwidth]{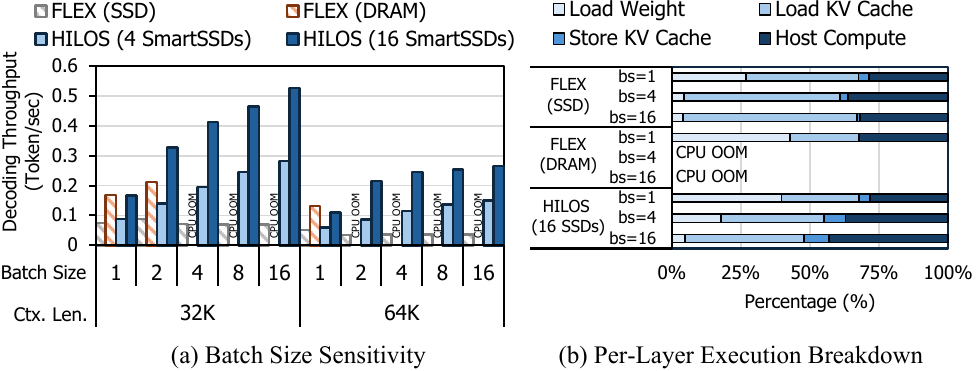}       
    \caption{
    Sensitivity to batch size with 66B model.  
    (a) Decoding throughput.  
    (b) Per-layer execution breakdown.
    }
    \label{fig:batch_sensi}
\end{figure}

\label{sec:batch_sensi}
\textbf{Batch Size Sensitivity.}
\cref{fig:batch_sensi}(a) presents a sensitivity analysis 
across various batch sizes.
\texttt{FLEX(DRAM)} is limited to a maximum batch size of two due to host DRAM capacity constraints.
Consequently, it exhibits a large \texttt{Load Weight} overhead, as detailed in \cref{fig:batch_sensi}(b).
While \texttt{FLEX(SSD)} supports larger batch sizes, its performance becomes bottlenecked by the KV cache I/O.
In contrast, \thiswork scales effectively up to a batch size of 16.
By mitigating both the weight-transfer and KV-cache overheads, \thiswork achieves superior throughput. 

\begin{figure}[t]
    \centering
    \Description{Figure.} 

    \includegraphics[width=\columnwidth]{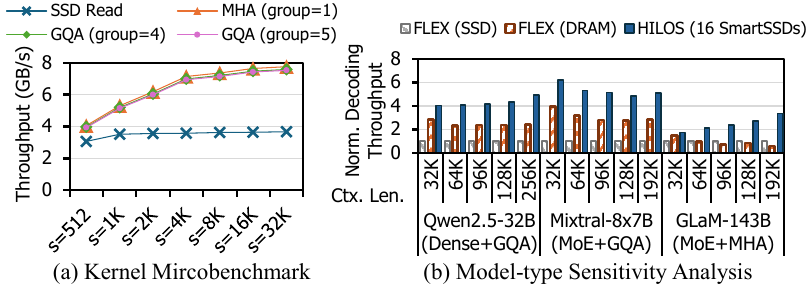}       
    \caption{
    Model architecture sensitivity analysis.
    }
    \label{fig:model_sensi}
\end{figure}

\label{sec:model_sensi}
\textbf{Sensitivity to Model Architectures.}
\cref{fig:model_sensi} evaluates the performance across diverse models.
\cref{fig:model_sensi}(a) shows the throughput 
of our attention kernels (\cref{tab:imple_result}) for KV data fetched from storage.
Although the GQA kernels exhibit a slightly lower throughput than $d_{group}=1$ kernel due to a higher arithmetic intensity, 
all kernels deliver far more than 3.0 GB/s, well exceeding the SSD's P2P read bandwidth.

\rev{ \cref{fig:model_sensi}(b) extends this analysis to end-to-end decoding throughput across context lengths, extending to the point of GPU memory saturation. 
\thiswork achieves 1.16$\times$ to 3.36$\times$ speedups over the baselines for models incorporating GQA or MoE.
While the lower KV-to-weight ratio of MoE and GQA brings slight favor to \texttt{FLEX(DRAM)}, \thiswork still outperforms baselines because it alleviates the significant memory pressure from longer contexts.
The performance gap widens as the context length increases because the baseline must limit the batch size to accommodate the growing memory footprint.
Furthermore, the RoPE recomputation overhead in \xcache remains negligible due to its efficient caching strategy~\cite{transformer_xl}. 
Thus, \thiswork demonstrates robust performance gains across diverse model architectures.
}

\begin{figure}[t]
    \centering
    \Description{Figure.} 
    \includegraphics[width=\linewidth]{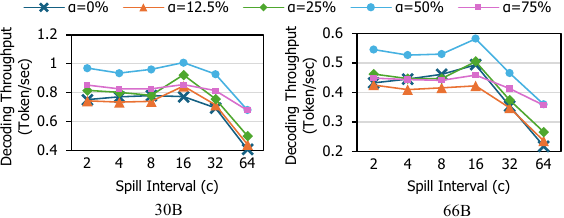}       

    \caption{
    Sensitivity study with varying spill intervals ($c$) across different X-Cache ratios ($\alpha$) and model sizes.  
    }
    \label{fig:c_sensi}
\end{figure}

\label{sec:c_sensi}
\textbf{Sensitivity to System Parameters.}
We evaluate \thiswork's performance sensitivity to the spill interval $c$ and the \xcache ratio $\alpha$.
Our profiling indicates an approximate bandwidth ratio of $B_{SSD}/B_{PCI}\approx 3$, where our analytical model predicts an optimal $\alpha\approx 50\%$.
\cref{fig:c_sensi} validates this analysis, as $\alpha=50\%$ consistently achieves the highest throughput.
The spill interval of $c=16$ performs best for all values of $\alpha$.
This choice aligns with the SSD's 4-KiB page granularity. 


\begin{figure}[t]
    \centering
    \Description{Figure.} 

    \includegraphics[width=\columnwidth]{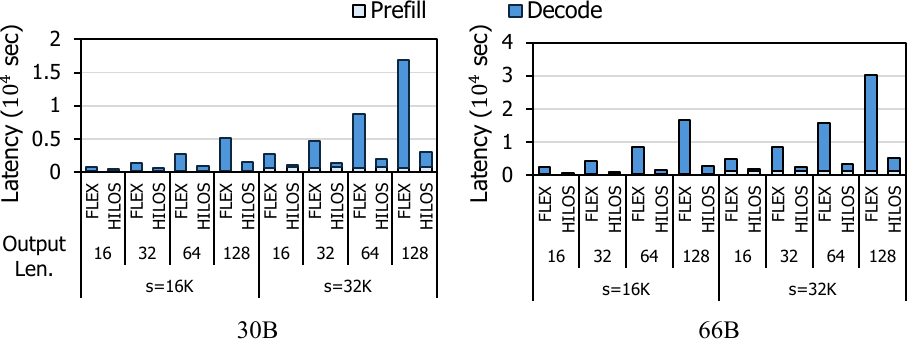}       
    \caption{
    Breakdown of total execution time by output length, model size ($B$), and context length ($s$).
    }
    \label{fig:output_sensi}
\end{figure}

\textbf{Output Sequence Length.}
\cref{fig:output_sensi} shows that longer output sequences yield greater speedups, reaching up to 6.08$\times$ over the baseline.
This is because the fixed prefill latency is amortized with longer outputs. 
Because the \wb mechanism leverages a write-back interval, $c$, 
efficient scaling for long sequences is ensured.

\subsection{Ablation Study Results}
\label{exp:ablation}
\begin{figure}[t]
    \centering
    \Description{Figure.} 

    \includegraphics[width=\columnwidth]{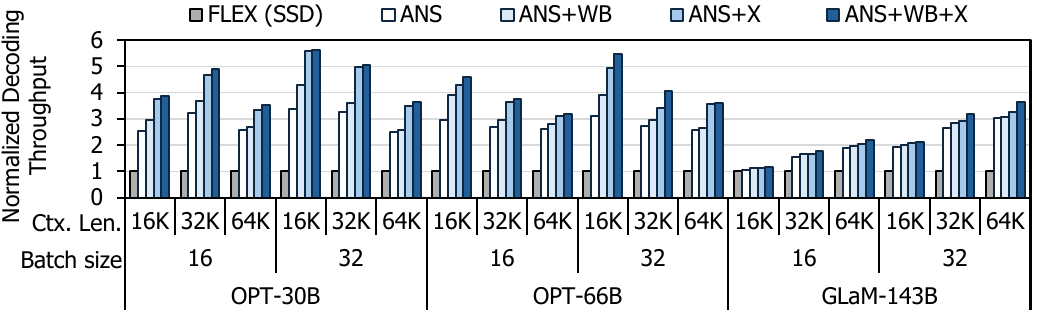}       
    \caption{
    An ablation study of \thiswork. 
    }
    \label{fig:ablation}
\end{figure}

\cref{fig:ablation} presents an ablation study of our optimizations, with throughput normalized to the \texttt{FLEX(SSD)} baseline.
Our \texttt{ANS} configuration, which excludes the write-back (\wb) and \xcache optimizations, provides up to a $3.39\times$ speedup.
Enabling the \wb mechanism (\texttt{ANS+WB}) yields an additional speedup of up to $1.32\times$ over \texttt{ANS} by hiding storage write latency.
The \xcache optimization (\texttt{ANS+X}) delivers a more substantial gain, achieving up to a $1.64\times$ speedup over \texttt{ANS}.
Although the overall improvements for the MoE model GLaM-143B are more modest due to its lower KV-cache-to-weight ratio, the relative benefits of our techniques scale more effectively with both longer context lengths and larger batch sizes.

\subsection{ Cost and Efficiency Analysis }
\label{exp:cost}


\begin{figure}[t]
    \centering
    \Description{Figure.} 

    \includegraphics[width=\columnwidth]{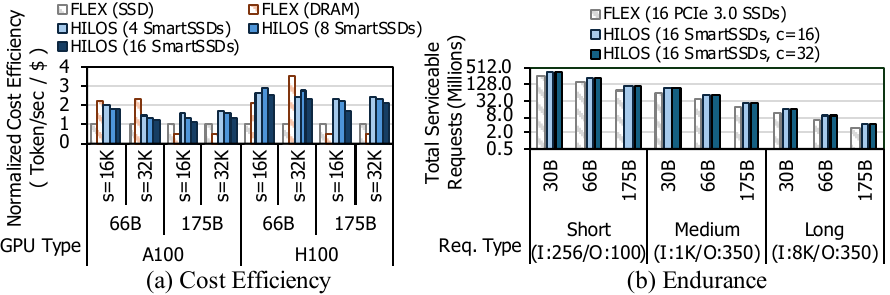}       
    \caption{
    (a) Cost-effectiveness and
    (b) endurance analysis.
    }

    \label{fig:cost}
\end{figure}

\textbf{Cost-Effectiveness.}
\cref{fig:cost}(a) presents a comprehensive analysis of the cost-effectiveness, measured in tokens per second per dollar (tokens/sec/\$), normalized to \texttt{FLEX(SSD)}.
The evaluation uses a baseline server with a \$15,000 host server and a \$7,000 A100 GPU.
The \thiswork configuration adds a \$10,000 PCIe expansion and sixteen SmartSSDs at \$2,400 each to this system, replacing conventional \$400 PCIe 4.0 SSDs.
For the 66B model, \thiswork achieves up to $2.02\times$ higher cost-efficiency than \texttt{FLEX(SSD)}.
While \texttt{FLEX(DRAM)} is $1.53\times$ more cost-effective when host DRAM is sufficient, its advantage diminishes for larger models.
For the 175B model, where DRAM capacity limits \texttt{FLEX(DRAM)}, \thiswork improves cost-efficiency by up to $1.68\times$.
In comparison, upgrading the GPU to a \$30,000 H100 provides a $1.39\times$ speedup, but I/O bottlenecks reduce its cost-efficiency. 
\thiswork delivers a comparable $1.29\times$ speedup while achieving $2.91\times$ greater cost-efficiency than the H100 configuration.

\label{exp:endurance}
\textbf{Endurance Analysis.}
Storage write endurance is often a cost concern, but \thiswork can avoid early SSD replacement due to two reasons.
First, the LLM inference workload exhibits a \emph{write-once, read-many} access pattern for the KV cache, where endurance is mainly limited by the total write volume.
Second, \thiswork employs two techniques to reduce total write traffic. 
The \xcache mechanism, with a cache rate of $\alpha$\%, lowers storage writes by approximately $\frac{\alpha}{2}$\%, since only half of the KV cache (X) needs to be stored.
Furthermore, \wb defers writes for newly generated KV cache entries, reducing write amplification.

\cref{fig:cost}(b) presents our endurance evaluation.
We classify requests as Small (I:256/O:100), Medium (I:1K/O:350), and Long (I:8K/O:350) based on Azure workload statistics~\cite{dynamo_llm}.
We estimate the total serviceable requests with 16 SmartSSDs, where each 3.84 TB SSD supports 7.008 PBW with a 3-month data retention~\cite{smartssd}.
Compared to the baseline, \thiswork improves endurance by 1.34$\times$ to 1.47$\times$.
Increasing the spill interval \texttt{c} from 16 to 32 yields an additional 1.02$\times$ to 1.05$\times$ improvement.
Even for long requests with the 175B model, our system supports over 4.08 million requests.

\begin{figure}[t]
    \centering
    \Description{Figure.} 

    \includegraphics[width=\columnwidth]{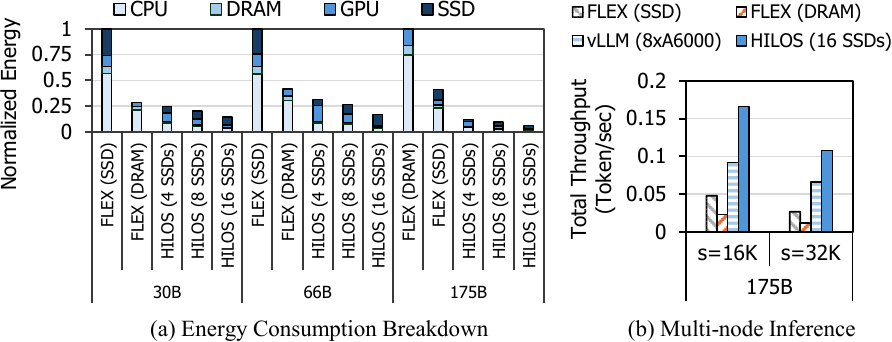}       
    \caption{
    (a) Energy consumption breakdown analysis. (b) Comparison with multi-GPU solutions using vLLM~\cite{vllm}.
    }
    \label{fig:cost2}
\end{figure}

\label{exp:energy}
\begin{sloppypar}
\textbf{Energy Efficiency.}
\cref{fig:cost2}(a) presents the energy consumption breakdown.
We measure GPU power using NVML~\cite{nvml} and CPU/DRAM power using Intel RAPL~\cite{rapl}.
The power for SmartSSDs is measured from  
its PCIe expansion board controller~\cite{aspeed} feature~\cite{falcon}.
For the baseline PCIe 4.0 SSD, we use its datasheet value of $13~W$~\cite{pm9a3}.
Although \thiswork's SmartSSDs consume more power than conventional SSDs, the significant reduction in inference latency yields superior energy efficiency.
The \texttt{FLEX(SSD)} baseline exhibits poor efficiency due to low throughput. 
While \texttt{FLEX(DRAM)} is efficient for the 30B and 66B models, its efficiency degrades for the 175B model due to batch size limitations.
\end{sloppypar}

\textbf{Multi-Node Solutions.}
In \cref{fig:cost2}(b), we compared \thiswork with a distributed multi-GPU configuration employing vLLM 0.9.1, which incorporates paged attention~\cite{vllm} and FlashAttention~\cite{flash_attn2}.
The system consists of two nodes utilizing tensor parallelism~\cite{shoeybi2019megatron} within each node and pipeline parallelism~\cite{gpipe, pipebd} across them.
Each node is equipped with four 48 GB RTX A6000 GPUs, 512 GB of host memory, an AMD EPYC 7302 CPU, and an InfiniBand EDR interconnect.
\thiswork achieves a 1.64$\times$ to 1.81$\times$ speedup over this multi-node baseline.
The distributed setup is bottlenecked by small batches and inter-node communication overheads, despite having a large aggregated GPU memory of 384 GB (8$\times$48GB).
While techniques like context parallelism~\cite{ring_attn, blockwise_attn} can efficiently parallelize long sequences, they do not solve the fundamental problem of fitting the large model 
into GPU memory. 
Consequently, supporting both long contexts and $>$100B parameter LLMs with context and model parallelism requires additional servers, which increases inter-server communication overhead, deployment cost~\cite{llmcost}, and power~\cite{dynamo_llm}.

\section{Discussion}
\label{sec:discussion}

\subsection{Applicability to ISP Solutions.}
\begin{figure}[t]
    \centering
    \Description{Figure.} 

    \includegraphics[width=\columnwidth]{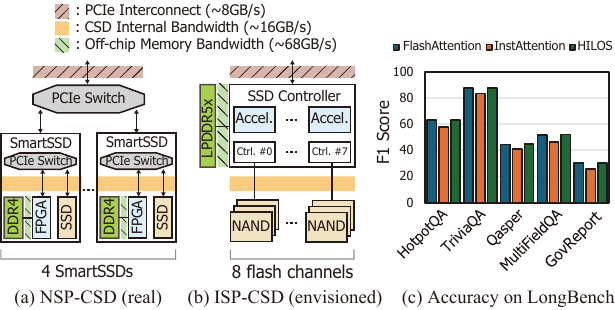}       
    \caption{
    (a) Our NSP-CSD prototype, which effectively emulates (b) an envisioned ISP-CSD.
    (c) F1 on LongBench~\cite{longbench}.
    }
    \label{fig:discussion}
\end{figure}

In this paper, we developed \thiswork on an NSP environment~\cite{smartssd}.
Although this is a solution that can be used today or in the near future, it has a limitation in that a single NSP device does not increase the internal bandwidth over a conventional system.
Fortunately, we envision that the core techniques used in \thiswork can also be applied to future In-Storage Processing (ISP) solutions~\cite{instattn, genstore,beacongnn,cambriconllm,megis}, which provide considerable bandwidth improvements with a single device.
\cref{fig:discussion}(b), illustrates an envisioned \thiswork-enabled ISP device, 
modeled after modern SSD designs~\cite{pm9a3, t500, 990pro}. 
The NAND flash array provides an aggregated capacity of 16~TB and is connected to the SSD controller via eight flash channels, each providing 2,000MT/s (total of 16~GB/s).
The single-package LPDDR5X (with four 16GB channels) as an internal DRAM offers an aggregated bandwidth of 68 GB/s.

From this, we argue that the performance benefit of this single PCIe 4.0 ISP unit would closely match that of the four SmartSSDs we used, in terms of the bandwidths of the internal storage (16 GB/s for 16 internal PCIe 3.0 lanes), system interconnect (8 GB/s for four PCIe 4.0 lanes), and the internal memory (52 GB/s for four DDR4 channels~\cite{smartssd}). 
%
Even though SmartSSDs are unable to fully emulate the envisioned ISP devices due to restricted firmware access, we believe they provide a reasonable proxy, 
given highly regular KV access patterns simplify SSD-internal managements~\cite{megis,genstore,ftl_page}. 


We evaluate the overhead of our proposed accelerator for ISP setup by synthesizing its design with $d_{group}=1$ using the OpenROAD Flow~\cite{openroad_dac, openroad_gomactech}, an open-source RTL-to-GDSII toolchain with the 45nm Nangate45 (FreePDK45) and scaled to an 8 nm process node to reflect modern SSD controllers' technology node~\cite{pm9a3} clocked at 300MHz (to match our FPGA configuration).
Along with the on-chip SRAM modeled using CACTI 7.0~\cite{cacti}, 
the total chip area is 0.47 mm$^2$, with power at 1.13 W on a 32K-token inference profile, which we believe is a reasonable overhead for ISP.
\textbf{Attention Offloading in ISP.}
The concept of offloading attention operations to ISP is explored in InstAttention~\cite{instattn}.
However, to meet the resource constraints of ISP, it adopts sparse KV retrieval techniques~\cite{sparq,instattn,infinigen,quest,omnicache,infinigen,h2o} with lossy compression.
While this method preserves accuracy for short context models~\cite{opt,llama2} up to 4K tokens, we observe noticeable accuracy drops for modern models with long contexts exceeding 32K tokens~\cite{llama3,llama4,qwen2,gemini15}.
To quantify this degradation, we evaluated Qwen2.5-32B-32K~\cite{qwen2} on five LongBench~\cite{longbench} datasets in \cref{fig:discussion}(c).
With InstAttention's default 1/8 compression ratio, the accuracy degrades by $3.52\sim5.73\%p$.
Unlike lossy schemes, the \thiswork accelerator (\cref{sec:arch}) maintains lossless accuracy compared to the widely used FlashAttention~\cite{flash_attn2}.

It is also worth noting that while InstAttention's attention offloading provides valuable insights, its practicality remains limited due to its implementation on a relatively outdated OpenSSD research platform~\cite{openssd1, openssd2} with emulation-based evaluation~\cite{nvmevirt} for multi-SSD configurations. 
In contrast, \thiswork is thoroughly studied on a modern hardware platform and addresses the practical issues that arise when attention offloading is employed in real NSP-enabled systems.

\rev{
\subsection{Future Computational Storage Device Designs} 
\label{sec:future}

Despite the architectural optimizations in \cref{sec:arch} and \cref{sec:fpga_details}, there remain several limitations posed by the existing CSD platforms. 
In specific, while our current CSD design saturates the bandwidth of PCIe 3.0 SSDs, future PCIe 5.0 devices will require a fourfold increase in throughput to fully utilize storage I/O, which would make it more challenging for CSD-based systems to maintain speedups.
In such regard,
we analyze current hardware limitations and explore architectural refinements to meet future demands.

First, softmax operations dominate execution time as the attention group size ($d_{group}$) increases.
It accounts for over 50\% of total execution time and consumes the majority of DSP resources due to computationally intensive floating-point exponential operations.
To match a 4$\times$ throughput increase from the assumed PCIe 5.0 interface via DSP parallelization, the design would require over 2,000 DSPs, which exceeds the total capacity of current SmartSSD platforms.
One solution we believe to be reasonable is to provide dedicated units for exponential functions. 
Given the prevalence of exponential functions in modern AI workloads such as sigmoid and GeLU, dedicated hardware units for them would significantly enhance the viability of CSDs for deep learning. 

Second, the adoption of separate clock domains with HLS would greatly facilitate timing closure.
Because contemporary HLS tools~\cite{vitis_hls} enforce a single clock domain, they inherently limit the scope of advanced frequency optimization.
Supporting distinct domains would allow the compute-intensive softmax logic to operate at high frequencies while maintaining lower frequencies for memory-bound GEMV logic.
This decoupling would provide a favorable trade-off between throughput and the strict power constraints of NSP.

Finally, the multi-SSD distributed execution model of \thiswork indicate that there might be a mismatch between existing hardware resources and actual operational requirements.
For instance, while KV Caches are partitioned across multiple CSDs to maximize aggregate internal bandwidth, this distribution results in a low storage footprint on each individual device.
Even under peak workloads, storage capacity overhead per SSD remains below 600 GB, leaving the 4 TB capacity of current SmartSSDs largely unexploited.
Furthermore, while current CSDs match internal and external lane bandwidths (\cref{fig:discussion}(a)), most traffic remains internal, leaving host-to-CSD lanes underutilized.
Higher internal bandwidth could be achieved through lightweight SSD mechanisms, such as coarse-grained block-level FTL mappings instead of DRAM-intensive page-level mappings \cite{megis, genstore, ftl_page}.
This approach is particularly effective because our design ensures sequential KV Cache accesses for both reads and writes through the write-back mechanism in \cref{sec:wb}.
To this extent, a more balanced design could be viable,
such as less capacity, more internal bandwidth, and/or more computational power per CSD. 
We believe such modifications would lead to a more cost-efficient system.
}

\subsection{Applicability to CXL-based Architectures}
The low-latency coherent memory access provided by Compute Express Link (CXL) \cite{cxl} offers a significant potential to further reduce data management overhead in \thiswork.
Our current cache management (\cref{sec:wb}) employs host-side buffering followed by explicit transmission to the near-storage accelerator.
This overhead scales with the spill interval ($c$) as shown in \cref{fig:c_sensi}.
This bottleneck is expected to intensify as modern SSDs adopt larger internal page granularities like 16 KiB \cite{16kb}, which necessitates more efficient data handling.


Physical memory isolation in PCIe-based environments necessitates explicit DMA orchestration via Xilinx Runtime (XRT) between FPGA DRAM and host DRAM.
This results in substantial software overhead and synchronization complexity~\cite{dongle}, reducing throughput by over 30\% when scaling $c$ from 4 KiB ($c$=16) to 16 KiB ($c$=64), as illustrated in \cref{fig:c_sensi}.
CXL.mem can address these limitations by enabling a unified address space, which eliminates explicit data copies and DMA management, thereby reducing latency and simplifying the underlying cache management logic.

\section{Related Work}
\label{sec:related}

\subsection{LLM Inference Acceleration.}
Prior work on LLM inference include kernel optimizations~\cite{deepspeedinf, flashllm}, parameter quantization~\cite{kvquant, atom, llm-qat, llmint8}, parallelism~\cite{ring_attn, scaletrans}, batching strategies~\cite{orca, dvabatch}, disaggregation~\cite{splitwise, distserve}, and MoE considerations~\cite{moe_lightning}.
For limited GPU resources, many approaches adopt an offloading strategy for both training~\cite{zerooffload, zeroinfinity, flashneuron, smartinfinity} and inference~\cite{flexgen, llminaflash, leviathan, powerinfer, hetegen, instattn,cambriconllm}.
\thiswork shares a similar spirit but leverages NSP for offline batched inference across long sequences, targeting applications like large-scale information extraction~\cite{narayan2018don, pu2023summarization, chang2024booookscore, goyal2022news, pang-etal-2023-long} and benchmarking~\cite{liang2023holistic, NEURIPS2023_89e44582, NEURIPS2023_91f18a12, guo2023gpt4graph}.

\rev{\subsection{Scalable Memory and Access}}
\textbf{Memory Expansion.} Emerging interconnects like CXL~\cite{cxl,cxl1,cxl2} and NVLink-C2C~\cite{hopper, nvlinkc2c} address memory capacity limitations. However, DRAM costs remain significantly higher than flash (approximately \$3~\cite{dram_cost} vs \$0.1~\cite{pm9a3} per GB~\cite{reis}), and scaling DRAM to accommodate the growing KV cache demands of long-context ~\cite{gemini15,llama3,qwen2,deepseekv3,llama4} and multi-turn models~\cite{multi_turn1,multi_turn2} is challenging. 
\rev{Although CXL-based near-data processing~\cite{lia, cxl_anns, cxl_offload, lpddrpim} shares design philosophies with our approach, \thiswork specifically targets PCIe-based storage environments.
Nevertheless, our principles of minimizing host-device traffic, enabling cooperative host-device processing, and ensuring lightweight accelerator design remain applicable to the CXL-based systems.

\textbf{Direct Memory Access.}
To minimize data management overheads, orchestrated DMA techniques~\cite{bam, dongle, gmt, instattn} allow accelerators to initiate storage I/O directly.
This mitigates the inefficiency of host-managed schemes~\cite{gds, spin, nvmmu} where the CPU manages NVMe interactions.
\thiswork aligns with this design principle to alleviate CPU bottlenecks.
Specifically, we offload the memory-intensive multi-head attention tasks to the near-storage accelerator, thereby effectively eliminating CPU-bound performance limitations.
}

\subsection{Near-Data Processing}
\textbf{Processing In Memory (PIM).}
While PIM architectures typically integrate computation within DRAM~\cite{newton, aim, hbmpim} and have seen commercial realization~\cite{upmem, hbmpim, aim}, recent works primarily apply DRAM-PIM~\cite{aim, hbmpim, piccolo, fala, gradpim} to LLM inference~\cite{transpim, pimdl, neupims, attacc, duplex, lpddrpim, paise, facil}.

\rev{
\textbf{Near-Storage Processing (NSP).}
Early NSP research focused on ``active'' disks~\cite{active_disk, active_disk_large, active_storage_large, idisk}, followed by embedded CPUs~\cite{self_sorting_ssd, earlysmartssd} or integrated processors~\cite{smartssd_datamine, biscuit}. 
Limitations in compute power led to the adoption of ASICs~\cite{genstore, deepstore, inspire} and FPGAs~\cite{netezza, exadata, rmssd, extrav, secndp}, eventually resulting in commercial platforms~\cite{smartssd} applied to diverse domains~\cite{smartssd_query, ann_smartssd, smartsage, smartinfinity, smartssd_dlrm}.
We distinguish \thiswork from prior NSP efforts that target FFN offloading~\cite{leviathan} or optimize for on-device low-batch scenarios~\cite{cambriconllm, lincoln, aif}.
Specifically, \thiswork targets KV-cache I/O to maximize throughput for batched offline inference.
By addressing DRAM capacity bottlenecks unique to this scenario, \thiswork achieves improved system-level throughput compared to prior architectures.
}

\section{Conclusion}
\label{sec:conclusion}

We present \thiswork, an NSP solution that accelerates offline LLM inference by offloading KV cache-related operations to near-storage accelerators.
By mitigating critical system-level integration challenges and employing a specialized hardware design, 
\thiswork achieves up to 7.86$\times$ higher throughput and 85\% energy reduction over state-of-the-art baselines in real system evaluations using a PyTorch-based inference pipeline.


\begin{acks}
\begin{sloppypar}
This work was supported by Samsung Memory Research Center (SMRC), National Research Foundation of Korea (NRF) grant  funded by the Korea government (MSIT) 
(2022R1C1C1011307, 
RS-2025-00519994), 
and
Institute of Information \& communications Technology Planning \& Evaluation (IITP) (RS-2024-00395134, 
RS-2024-00347394, 
RS-2023-00256081, 
RS-2021-II211343,
RS-2024-00459026 
).
Part of the infrastructure used in this work was supported by Korea Basic Science Institute (National research Facilities and Equipment Center) grant funded by the Ministry of Science and ICT  
(No. RS-2025-00564840).
Jinho Lee is the corresponding author.
\end{sloppypar}
\end{acks}


\appendix
\section{Artifact Appendix}

\subsection{Abstract}

This artifact provides the source code and hardware configuration for \thiswork, a high-throughput NSP system optimized for long-context LLM offline inference.
It comprises: (1) HLS source code for FPGA-based attention acceleration kernels and (2) host-side system software.
The artifact supports both the end-to-end synthesis of FPGA binaries and the deployment of the full inference pipeline.

\subsection{Description}

\subsubsection{How to Access}

The source code is permanently archived via Zenodo at \url{https://doi.org/10.5281/zenodo.18162119}.
The latest development version is maintained at \url{https://github.com/hongsunjang/HILOS/tree/asplos26}.


\subsubsection{Hardware dependencies}
\label{ae:hw_dep}
Reproducing the experiments requires the following hardware configuration.
\begin{itemize}
    \item \textbf{GPU:} At least one NVIDIA GPU with GPUDirect Storage (GDS) support.
    \item \textbf{Storage:} At least one Samsung SmartSSD.
    \item \textbf{CPU:} An Intel CPU is mandatory for SmartSSD compatibility.
\end{itemize}

\subsubsection{Software dependencies}
\label{ae:sw_dep}
The environment relies on the following components.

\begin{itemize}
    \item \textbf{Drivers and CUDA:} NVIDIA driver (version 530 or higher) and CUDA/CuDNN toolkit (version 12.1).
    \item \textbf{GPUDirect Storage (GDS):} The NVIDIA GDS package appropriate for the system kernel.
    \item \textbf{FPGA Tools:} Xilinx Vivado and Vitis HLS (version 2023.1).
    \item \textbf{Xilinx Runtime (XRT):} The XRT version matching the target SmartSSD platform. A cold reboot is mandatory after installation to flash the SmartSSD.
\end{itemize}

Source the settings scripts prior to execution.
\begin{verbatim}
$ source <path_to_Vitis_HLS>/2023.1/settings64.sh
$ source /opt/xilinx/xrt/setup.sh
\end{verbatim}

\subsection{Installation}
\label{ae:install}
For detailed, step-by-step instructions on environment setup and dependency installation, please refer to the \texttt{README.md} file in the root directory of the repository:
\url{https://github.com/hongsunjang/HILOS/tree/asplos26}

\subsection{Experiment Workflow}
\label{ae:workflow}

The evaluation follows a two-step workflow involving (1) FPGA binary synthesis and (2) LLM inference deployment.

\subsubsection{Step 1: HLS Synthesis and Verification}

\begin{enumerate}
    \item \textbf{Verify HLS Functionality:} Run the test script to validate the HLS implementation.
    \begin{verbatim}
$ python tests/test_llm.py --mode hls_gqa
    \end{verbatim}

    \item \textbf{Verify Binary Synthesis:} Copy the generated bitstream to the \texttt{bins} folder and verify functionality.
    \begin{verbatim}
$ python tests/test_llm.py --mode fpga_gqa
    \end{verbatim}
    Verify that the token output matches the expected values and that the kernel executes without errors.
\end{enumerate}

\subsubsection{Step 2: LLM Inference Deployment}
\mbox{}

\begin{enumerate}
    \item \textbf{Proposed Method (ANS):} Execute the system with FPGA acceleration.
    \begin{verbatim}
$ python3 bench_suite.py hilos
    \end{verbatim}
    \item \textbf{Proposed Method (+X-Cache):} Execute \thiswork combined with the X-Cache optimization.
    \begin{verbatim}
$ python3 bench_suite.py xcache
    \end{verbatim}
\end{enumerate}





\bibliographystyle{unsrt} 
\balance
\bibliography{refs}

\end{document}